\documentclass{article}

\usepackage{microtype}
\usepackage{graphicx}
\usepackage{subfigure}
\usepackage{booktabs} 
\usepackage{enumitem}

\usepackage{hyperref}

\usepackage[accepted]{icml2025}

\usepackage{amsmath}
\usepackage{amssymb}
\usepackage{mathtools}
\usepackage{amsthm}

\usepackage[capitalize,noabbrev]{cleveref}

\theoremstyle{plain}

\theoremstyle{definition}

\theoremstyle{remark}

\usepackage[textsize=tiny]{todonotes}

\icmltitlerunning{The AI Agent Index}

\begin{document}

\twocolumn[
\icmltitle{The AI Agent Index}



\icmlsetsymbol{equal}{*}

\begin{icmlauthorlist}
\icmlauthor{Stephen Casper}{equal,mit}
\icmlauthor{Luke Bailey}{stanford}
\icmlauthor{Rosco Hunter}{warwick}
\icmlauthor{Carson Ezell}{harvard}
\icmlauthor{Emma Cabalé}{saclay}
\icmlauthor{Michael Gerovitch}{mit}
\icmlauthor{Stewart Slocum}{mit}
\icmlauthor{Kevin Wei}{harvard}
\icmlauthor{Nikola Jurkovic}{harvard}
\icmlauthor{Ariba Khan}{mit}
\icmlauthor{Phillip Christoffersen}{mit}
\icmlauthor{A. Pinar Ozisik}{mit}
\icmlauthor{Rakshit Trivedi}{mit}
\icmlauthor{Dylan Hadfield-Menell}{mit}
\icmlauthor{Noam Kolt}{equal,hebrew}
\end{icmlauthorlist}

\icmlaffiliation{mit}{Massachusetts Institute of Technology}
\icmlaffiliation{stanford}{Stanford University}
\icmlaffiliation{warwick}{University of Warwick}
\icmlaffiliation{harvard}{Harvard University}
\icmlaffiliation{saclay}{École Normale Supérieure Paris-Saclay, Université Paris-Saclay}
\icmlaffiliation{hebrew}{Hebrew University}

\icmlcorrespondingauthor{Stephen Casper}{scasper@mit.edu}
\icmlcorrespondingauthor{Noam Kolt}{noam.kolt@mail.huji.ac.il}

\icmlkeywords{Machine Learning, ICML}

\vskip 0.3in
]



\printAffiliationsAndNotice{\icmlEqualContribution} 

\begin{abstract}
Leading AI developers and startups are increasingly deploying agentic AI systems that can plan and execute complex tasks with limited human involvement. 
However, there is currently no structured framework for documenting the technical components, intended uses, and safety features of agentic systems. 
To fill this gap, we introduce the \textbf{AI Agent Index}, \textit{the first public database to document information about currently deployed agentic AI systems}.
For each system that meets the criteria for inclusion in the index, we document the system’s components (e.g., base model, reasoning implementation, tool use), application domains (e.g., computer use, software engineering), and risk management practices (e.g., evaluation results, guardrails), based on publicly available information and correspondence with developers. 
We find that while developers generally provide ample information regarding the capabilities and applications of agentic systems, they currently provide limited information regarding safety and risk management practices.
The AI Agent Index is available online at
\href{https://aiagentindex.mit.edu/}{https://aiagentindex.mit.edu/},
with raw data 
at \href{https://docs.google.com/spreadsheets/d/14O8k6ttvM-Zgp5aIdmxvP-KjsUy99O23r0LDwQJOh_g/edit?usp=sharing}{this link}.
\end{abstract}

\section{Introduction} \label{sec:intro}

`Agentic' AI systems that can be instructed to plan and directly execute complex tasks with only limited human involvement \citep{xi2023rise, wang2024survey, durante2024agent, sager2025ai} are transitioning from research prototypes to real-world products (e.g., 
\href{https://devin.ai/}{Devin}, 
\href{https://h2o.ai/platform/enterprise-h2ogpte/}{h2oGPTe}, 
\href{https://usesimple.ai/}{Simple AI}, 
\href{https://xbow.com/}{XBOW}). 
These systems---which are generally comprised of foundation
models augmented with scaffolding for reasoning,
planning, memory, and tool use \citep{sumers2023cognitive, compound-ai-blog, yao2024language, su-etal-2024-language}---are being deployed in a growing number of domains (see \Cref{fig:application_domains}).

\begin{figure*}[tp]
    \centering
    \begin{minipage}[t]{0.48\linewidth}
        \centering
        \includegraphics[width=\linewidth]{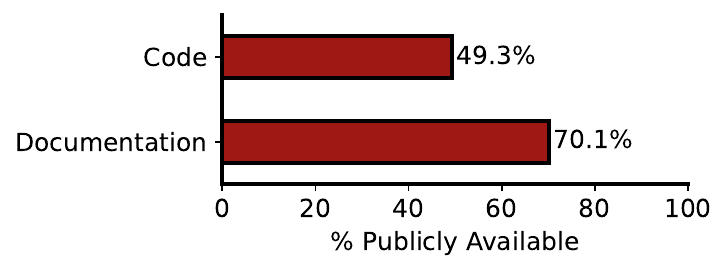}
        \caption{Most AI agent developers in the index provide some public documentation (70.1\%), while about half (49.3\%) release their underlying code.
        }
        \label{fig:accessibility}
    \end{minipage}
    \hfill
    \begin{minipage}[t]{0.48\linewidth}
        \centering
        \includegraphics[width=\linewidth]{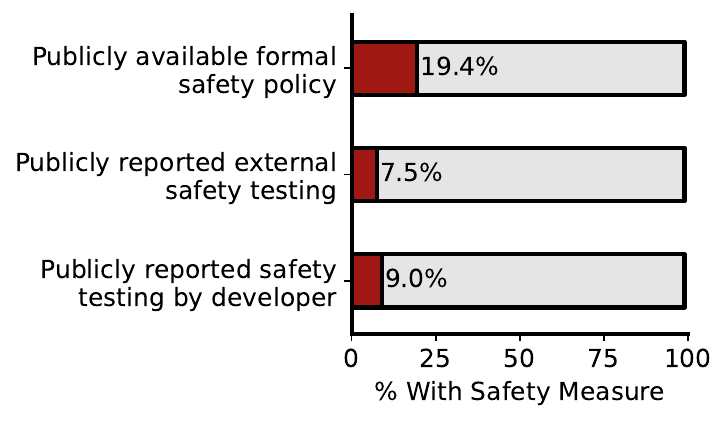}
        \caption{Only 19.4\% of indexed agentic systems disclose a formal safety policy, and fewer than 10\% report external safety evaluations.
        }
        \label{fig:safety}
    \end{minipage}
\end{figure*}

The performance of agentic systems is steadily improving on benchmarks \citep{mialon2023gaia, xie2024osworld, zhou2023webarena, koh2024visualwebarena, yoran2024assistantbench, xu2024theagentcompany}, and these systems are being integrated into broader swathes of economic activity \citep{wang2024survey, durante2024agent, sager2025ai}. As a result, their real-world impacts are mounting  \citep{chan2023harms, gabriel2024ethics, anwar2024foundational, kolt2025governing}. 
Alongside the significant opportunities presented by agentic systems, researchers have also raised noteworthy concerns, including cybersecurity risks \citep{fang2024llm, fang2024teams}, loss of control \citep{cohen2024regulating, bengio2025internationalaisafetyreport}, and physical harm where agents operate robotic systems \citep{ruan2023identifying}.

Despite growing efforts to study trends in the development of agentic AI systems, including evaluating their performance and cost \citep{kapoor2024ai, hal}, assessing their potential harms \citep{andriushchenko2024agentharm, kumar2024refusal, US_AISI_2025}, and increasing visibility into their operation \citep{shavit2023practices, chan2024visibility, chan2024ids, chan2025infrastructureaiagents, kolt2025governing}, many practical questions remain unanswered: 

\begin{itemize}[itemsep=4pt, topsep=2pt]
    \item Which organizations are developing agentic systems?
    \item In which domains are they being deployed?
    \item What infrastructure do agentic systems require?
    \item How is their performance and safety evaluated?
    \item What guardrails are used to mitigate risks?
\end{itemize}

To empirically answer these questions and improve public understanding of agentic AI systems, we introduce and release the \textbf{AI Agent Index}, a comprehensive sample of deployed agentic AI systems (n = 67). 
The index, which is constructed from a combination of publicly available data and correspondence with developers, documents publicly-available information on the intended uses of agentic systems, their technical components (including reasoning, planning, and memory implementation, base models, observation and action space, and user interface), safety features (including accessibility of system components, usage controls and restrictions, and red-teaming practices), and details regarding the organizations developing and deploying agentic systems (including entity type and country of origin). 

In addition to collecting and systematizing information about agentic AI systems, the index also sheds light on the \textit{availability} of such information. 
Specifically, we find that while relatively detailed information is available regarding the features and applications of agentic systems (\Cref{fig:accessibility}), strikingly limited information is available regarding their safety evaluations and guardrails (\Cref{fig:safety}).

In this paper, we make three contributions: 
\begin{enumerate}
    \item We introduce a structured framework for documenting the technical, safety, and policy-relevant features of agentic AI systems.
    \item We identify currently deployed agentic systems that meet our criteria (described below) and publicly document these systems according to our framework.
    \item We discuss key findings from the index, shedding light on geographic spread, academic vs. industry development, openness, and risk management of agentic systems.
\end{enumerate}

The index is available on the web at
\href{https://aiagentindex.mit.edu/}{https://aiagentindex.mit.edu/} with raw data accessible \href{https://docs.google.com/spreadsheets/d/14O8k6ttvM-Zgp5aIdmxvP-KjsUy99O23r0LDwQJOh_g/edit?usp=sharing}{here}.

\section{Background}
\label{sec:background}

\textbf{There is no widely accepted definition of ``AI agent''.} 
The notion of artificial agency has a long and contentious history, spanning multiple decades and diverse disciplines.
These include cybernetics \cite{rosenblueth1943behavior, ashby1956introduction, wiener1961cybernetics}, artificial life \citep{maes1990designing, maes1993modeling, maes1995artificial}, rational agency \citep{rao1991modeling}, software engineering \citep{wooldridge1995intelligent, jennings2000agent}, reinforcement learning \citep{sutton2018reinforcement}, and philosophy \citep{dennett1989intentional, dung2024understanding}.
While there have been notable attempts to define the term ``agent'', including in the context of computational systems \citep{franklin1996agent,russell2020, kenton2023discovering}, we do not decide among these definitions or offer an alternative definition. 
Instead, we follow \citet{chan2023harms}, and loosely characterize agentic AI systems as ones that exhibit, to some significant degree, a combination of the following properties: 

\begin{enumerate} [label=\alph*), itemsep=0pt, topsep=0pt]
    \item \textbf{Underspecification}: the system can accomplish a goal provided to it without a precise specification of how to do so. 
    \item \textbf{Directness of impact}: the system’s actions can affect the world with little to no human mediation. 
    \item \textbf{Goal-directedness}: the system acts as if in the pursuit of a particular objective. 
    \item \textbf{Long-term planning:} the system can solve problems by reasoning about how to approach them, constructing plans, and executing them step by step.
\end{enumerate}

\subsection{Agentic Architectures, Applications, and Opportunities} \label{sec:architectures}

Contemporary AI agents are generally compound systems \citep{compound-ai-blog} comprised of a foundation model augmented by external resources, known as ``scaffolding'', which enable effective planning, memory, and tool use \citep{wang2024survey, xi2023rise, durante2024agent}. 
Planning of complex series of actions is typically facilitated through chain-of-thought-based reasoning processes \cite{wei2022chain, yao2022react, yao2023tree, shinn2023reflexion, openai2024o1preview}. 
Memory relies on information stored in the base model and/or in external storage modules \citep{sumers2023cognitive}. 
Tool use is enabled through API calls and natural language dialogue between the base model and external software, databases, and other affordances \citep{schick2023toolformer, mialon2023augmented, qin2023toolllm}.

These agentic architectures are increasingly applied to a variety of domains, including programming \citep{jimenez2023swe, yang2024swebenchmultimodal}, machine learning research \citep{huang2024mlagentbench,wijk2024re,chan2024mle}, experimentation in the natural sciences \citep{boiko2023autonomous, bran2024augmenting, jansen2024discoveryworld}, and consumer activities such as online retail \citep{yao2022webshop, deng2023mind2web}, travel planning \citep{xie2024travelplanner}, and general-purpose web browsing (\citealp{gur2023real,wu2024copilot}). 
Progress in these applications is being evaluated by a growing suite of benchmarks, which measure performance in computer use \citep{mialon2023gaia, xie2024osworld, zhou2023webarena, koh2024visualwebarena, yoran2024assistantbench}, software engineering \citep{jimenez2023swe, yang2024swebenchmultimodal}, and virtual work environments \cite{xu2024theagentcompany}.

\subsection{Safety Risks and Ethical Concerns} \label{sec:risks}

Given that agentic AI systems are built on foundation models, they are susceptible to many of the risks associated with such models, including harms arising from hallucinations, biased outputs, and leakage of private data \citep{bender2021dangers, weidinger2022taxonomy, solaiman2023evaluating}.
Agentic systems, however, also present new risks that stem specifically from their agentic properties, i.e., underspecification, directness of impact, goal-directedness, and long-term planning \citep{chan2023harms, cohen2024regulating, ruan2023identifying, andriushchenko2024agentharm, bengio2025internationalaisafetyreport}. 
For example, while chatbots often cause harm by human users acting upon model outputs (e.g., deploying model-generated malicious code) \citep{phuong2024evaluating}, agentic AI systems can \textit{directly} cause harm (e.g., autonomously hacking websites) \citep{fang2024llm, jaech2024openai}.

Additionally, as agentic AI systems undertake more complex and long-horizon tasks, with limited human oversight, users are likely to repose greater trust in those systems, potentially developing asymmetric relationships of dependence \citep{gabriel2024ethics, manzini2024should, manzini2024code, bengio2025internationalaisafetyreport}. 
Moreover, agentic systems developed and operated by large platform companies could enable those companies to exert greater influence and control over users and third parties with whom they interact (e.g., vendors accessed through platform-controlled agents) \citep{lazar2024frontier}.

\subsection{Documentation Frameworks} \label{sec:documentation}

Many frameworks have been developed to document the features of AI systems,  the resources used to build them, and the contexts in which they are deployed. These include datasheets \citep{gebru2018datasheets}, model cards \citep{mitchell2019model}, reward reports \citep{gilbert2022reward}, ecosystem graphs \citep{bommasani2023ecosystem}, and data provenance cards \citep{longpre2023data}.  
In addition, several databases have been created to collect information regarding contemporary AI systems and their real-world impacts, such as the Foundation Model Transparency Index \citep{bommasani2023foundation}, the AI Incident Database \citep{mcgregor2021preventing}, and the AI Risk Repository \citep{slattery2024ai}. 
Currently, however, there are no equivalent frameworks for documenting agentic AI systems. This lack of structured information limits both researchers' ability to study and build agentic systems, as well as policymakers' capacity to design appropriate governance mechanisms \citep{winecoff2024improving}.

The AI Agent Index fills this gap.
By collecting and communicating technical, safety, and policy-relevant information concerning agentic AI systems, the index aims to inform different stakeholders in distinct ways. Specifically, the index:

\begin{enumerate}
    \item Enables \textbf{users} to better understand the capabilities and limitations of agentic systems with which they interact.
    \item Provides \textbf{developers} more comprehensive and granular information about currently deployed agentic systems.
    \item Supports \textbf{auditors} and red-teams in deciding the scope and focus of their evaluations of agentic systems.
    \item Offers an evidence base to \textbf{policymakers} designing governance mechanisms for agentic systems.
    \item Improves \textbf{public} awareness and understanding of agentic systems.
\end{enumerate}

\section{Methods}
\label{sec:methods}

\textbf{What does the index include?} 
As discussed in \Cref{sec:background}, there is no widely-accepted definition of ``AI agent.'' 
We do not propose one here. 
Given our focus on the societal impacts of agentic AI systems, we draw on the four characteristics introduced by \citet{chan2023harms} discussed in \Cref{sec:background}.
Importantly, to address the practical questions outlined in \Cref{sec:intro}, we primarily document the features of agentic AI systems that are either deployed as products or available open source.  

\begin{figure*}[t!] 
    \centering
    \includegraphics[width=1.0\linewidth]{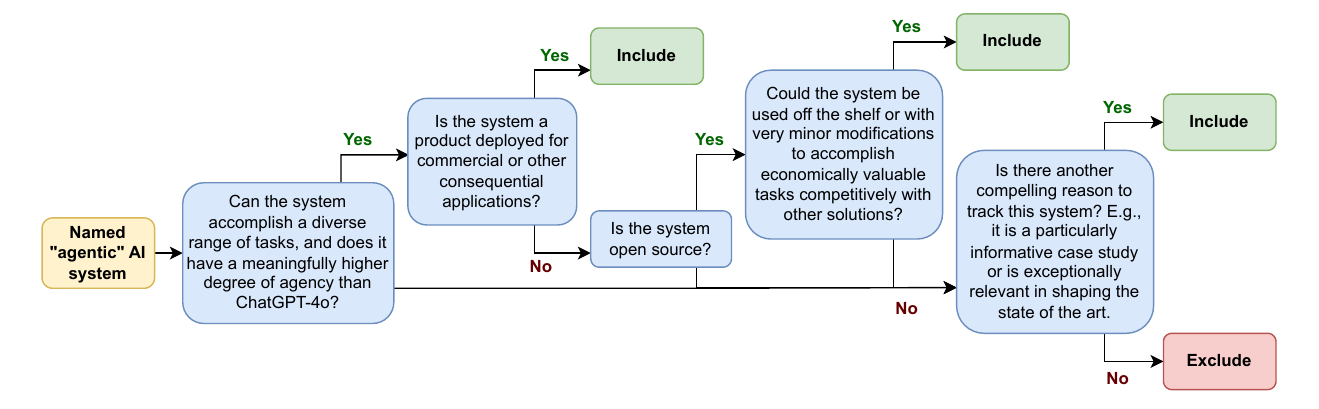}
    \caption{Decision graph for determining inclusion in the index: We focused on indexing agentic \emph{systems} (as opposed to models or development frameworks) and drew on the four characteristics of agency from \citet{chan2023harms}: \textit{underspecification}, \textit{directness of impact}, \textit{goal-directedness}, and \textit{long-term planning}. In total, we indexed 67 systems.}
    \label{fig:inclusion}
\end{figure*}

The full decision graph we used to determine inclusion in the index is shown in \Cref{fig:inclusion}.
Notably, we restricted the index to agentic \emph{systems} and did not include language models themselves, or agent development frameworks (unless the framework was built around a qualifying flagship system, in which case we indexed that system). 
We also created a single index entry per named and versioned system. 
Different releases (e.g., ``HelpfulAgent1.1'' vs ``HelpfulAgent1.2'') and different configurations (e.g., ``HelpfulAgent-Claude3.5-Sonnet'' vs. ``HelpfulAgent-GPT4o'') were indexed under the same entry. 
The final node in our decision graph (\Cref{fig:inclusion}) facilitates the inclusion of systems that otherwise would not strictly fit the criteria at our discretion. 
In practice, we only invoked this for systems from leading companies that were announced but have not (yet) been externally deployed, such as \href{https://www.youtube.com/watch?v=SKBG1sqdyIU&t=1s}{OpenAI o3} or \href{https://deepmind.google/technologies/project-mariner/}{Project Mariner}. 
In total, we indexed 67 systems.
Limitations of our methods are discussed in \Cref{sec:limitations}.

The AI Agent Index represents a snapshot in time as of \textbf{December 31, 2024}. 
New developments in the AI agent research and product ecosystem occur weekly. 
To improve thoroughness and consistency, we only indexed systems announced by, and available in, 2024.

\textbf{What does the index \textit{not} include?} 
Our selection criteria led us to exclude the following types of systems:
\begin{itemize}
    \item \textbf{Non-``agentic'' models} such as Llama-3.2-90B-Vision-Instruct \citep{dubey2024llama}.
    \item \textbf{Unnamed systems} often comprised of simple baseline implementations introduced under frameworks or benchmarks such as CORE-Bench \citep{Siegel2024COREBenchFT}, AgentHarm \citep{andriushchenko2024agentharm}, or The Agent Company \citep{xu2024theagentcompany}. 
    \item \textbf{Non-``agentic'' development frameworks without a qualifying flagship model} such as AutoGPT \citep{firat2023if}, \href{https://beam.ai/}{Beam}, \href{https://dust.tt/}{Dust}, \href{https://www.gumloop.com/}{GumLoop}, \href{https://www.lindy.ai/}{Lindy}, \href{https://github.com/openai/swarm}{OpenAI Swarm}, \href{https://qwen.readthedocs.io/en/latest/framework/qwen_agent.html}{Qwen-Agent}, or \href{https://spell.so/}{Spell}.
    \item \textbf{Systems that cannot open-endedly accomplish a diverse range of tasks} such as systems that propose solutions to git requests (e.g., \href{https://mentat.ai/blog/mentatbot-sota-coding-agent}{MentatBot}, \href{https://www.enginelabs.ai/}{Engine},  Globant Code-Fixer Agent \citep{globant_code_fixer_agent_2024}).
    \item \textbf{Systems that do not have a meaningfully higher degree of agency than ChatGPT-4o}\footnote{ChatGPT-4o allows users to customize system prompts, can engage in open-ended dialogue, and can search/synthesize web searches when responding to users.} (based on the four aspects of agency from \citet{chan2023harms}) such as \href{https://www.taskade.com/}{Taskade}, \href{https://www.vonage.com/unified-communications/features/ai-virtual-assistant/}{Vonage AI Virtual Assistant}, \href{https://www.talkdesk.com/}{Talkdesk}, \href{https://www.ibm.com/watsonx}{IBM WatsonX}, and \href{https://actionagents.co/}{ActionAgents}.
    \item \textbf{Systems that are not open source or products} deployed for commercial or other consequential applications such as Falcon-UI \citep{shen2024falcon} or HoneyComb \citep{zhang2024honeycomb}.
    \item \textbf{Open source systems that could not be used competitively off the shelf}, often due to age or narrow scope such as \href{https://genia-dev.github.io/GeniA/}{GeniA}, ReAct \citep{Yao2022ReActSR}, Pearl \citep{Sun2023PEARLPL}, or \href{https://github.com/aorwall/moatless-tools}{Moatlesss}.
    \item \textbf{Systems deployed after the cutoff date} of December 31, 2024 such as \href{https://api-docs.deepseek.com/news/news250120}{Deepseek-R1}, \href{https://team.doubao.com/zh/special/doubao_1_5_pro}{Doubao-1.5-pro}, or \href{https://openai.com/index/introducing-operator/}{OpenAI Operator}. 
\end{itemize}

\textbf{How was information collected?} From August 2024 to January 2025, we identified agentic AI systems using web searches, academic literature review, benchmark leaderboards (e.g., SWE-bench \citep{jimenez2023swe} and GAIA \citep{mialon2023gaia}), and additional resources that compile lists of agentic systems (e.g., \href{https://aiagentslist.com/}{https://aiagentslist.com/}, \href{https://vyokky.github.io/LLM-Brained-GUI-Agents-Survey/}{https://vyokky.github.io/LLM-Brained-GUI-Agents-Survey/}, and \href{https://www.letta.com/blog/ai-agents-stack}{https://www.letta.com/blog/ai-agents-stack}).

On a rolling basis, we created the first drafts of agent cards according to the template outlined next in \Cref{sec:agent_card_components}. 
After each first draft was completed, we contacted the developers of each agent to request feedback and potential corrections. 
We received a 36\% response rate. 
After editing each draft to incorporate feedback, we updated and finalized agent cards in January 2025 to ensure that all reflected the state of the field as of December 31, 2024. 
For all web sources cited in all agent cards (excluding stable papers, videos, and social media posts), we cited stable archived versions of websites preceding and as close to December 31, 2024 as possible using \href{https://web.archive.org/}{https://web.archive.org/} and \href{https://perma.cc/}{https://perma.cc/}.

\section{Agent Card Components}
\label{sec:agent_card_components}

Each agent card contains 33 fields of information, divided into 6 categories:
\begin{enumerate}

    \item \textbf{Basic information}
    \begin{itemize}
        \item Website
        \item Short description
        \item Intended uses: \textit{What does the developer state that the system is intended for?}
        \item Date(s) deployed
    \end{itemize}

    \item \textbf{Developer}
    \begin{itemize}
        \item Website
        \item Legal name
        \item Entity type
        \item Country (location of developer or first author's first affiliation)
        \item Safety policies: \textit{What safety and/or responsibility policies are in place?}
    \end{itemize}

    \item \textbf{System components}
    \begin{itemize}
        \item Backend model: \textit{What model(s) are used to power the system?}
        \item Publicly available model specification: \textit{Is there formal documentation on the system’s intended uses and how it is designed to behave in them?}
        \item Reasoning, planning, and memory implementation: \textit{How does the system ‘think’?}
        \item Observation space: \textit{What is the system able to observe while `thinking’?}
        \item Action space/tools: \textit{What direct actions can the system take?}
        \item User interface: \textit{How do users interact with the system?}
        \item Development cost and compute: \textit{What is known about the development costs?}
    \end{itemize}

    \item \textbf{Guardrails and oversight}
    \begin{itemize}
        \item Accessibility of components:
        \begin{itemize}
            \item Weights: \textit{Are model parameters available?}
            \item Data: \textit{Is data available?}
            \item Code: \textit{Is code available?}
            \item Scaffolding: \textit{Is system scaffolding available?}
            \item Documentation: \textit{Is documentation available?}
        \end{itemize}
        \item Controls and guardrails: \textit{What notable methods are used to protect against harmful actions?}
        \item Customer and usage restrictions: \textit{Are there know-your-customer measures or other restrictions on customers?}
        \item Monitoring and shutdown procedures: \textit{Are there any notable methods or protocols that allow for the system to be shut down if it is observed to behave harmfully?}
    \end{itemize}

    \item \textbf{Evaluations}
    \begin{itemize}
        \item Notable benchmark evaluations (e.g., on SWE-Bench Verified)
        \item Bespoke testing (e.g., demos)
        \item Safety: \textit{Have safety evaluations been conducted by the developers? What were the results?}
        \item Publicly reported external red-teaming or comparable auditing:
        \begin{itemize}
            \item Personnel: \textit{Who were the red-teamers/auditors?}
            \item Scope, scale, access, and methods: \textit{What access did red-teamers/auditors have and what actions did they take?}
            \item Findings: \textit{What did the red-teamers/auditors conclude?}
        \end{itemize}
    \end{itemize}

    \item \textbf{Ecosystem}
    \begin{itemize}
        \item Interoperability with other systems: \textit{What tools or integrations are available?}
        \item Usage statistics and patterns: \textit{Are there any notable observations about usage?}
    \end{itemize}

    \item \textbf{Additional notes:} If any
    
\end{enumerate}

We populated each field in each card with written notes based on publicly available information. 
When no information was available, we recorded ``None'' or ``Unknown.''

\section{Findings}
\label{sec:findings}

In addition to compiling specific information regarding each of the 67 indexed systems, the AI Agent Index offers a high-level perspective of this emerging field.
Noting the limitations and biases discussed next (in \Cref{sec:limitations}), here, we offer a bird's eye view of the state of the art for AI agents. 

\begin{figure}[h!]
    \centering
    \includegraphics[width=\linewidth]{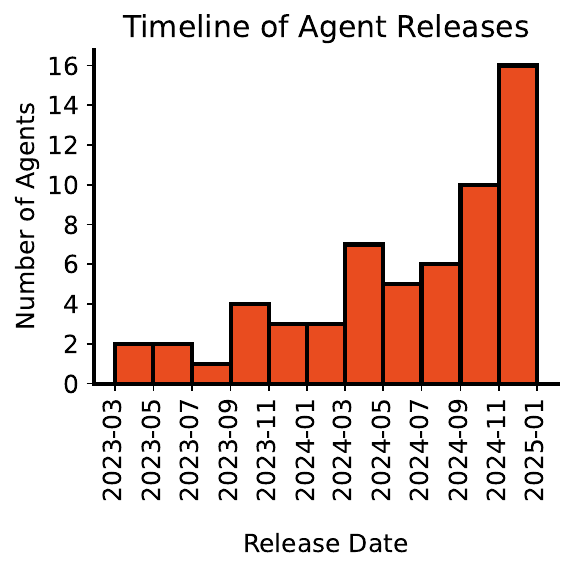}
    \caption{Agentic systems are being deployed at a steadily increasing rate.}
    \label{fig:timeline}
\end{figure}

\textbf{Agentic systems are being deployed at a steadily increasing rate.} 
Systems that meet our criteria for inclusion in the index have had (initial) deployments dating back to early 2023. 
However, \Cref{fig:timeline} shows that they have been deployed at an increasing rate with approximately half of the indexed systems deployed in the second half of 2024. 

\begin{figure}[h!]
    \centering
    \includegraphics[width=\linewidth]{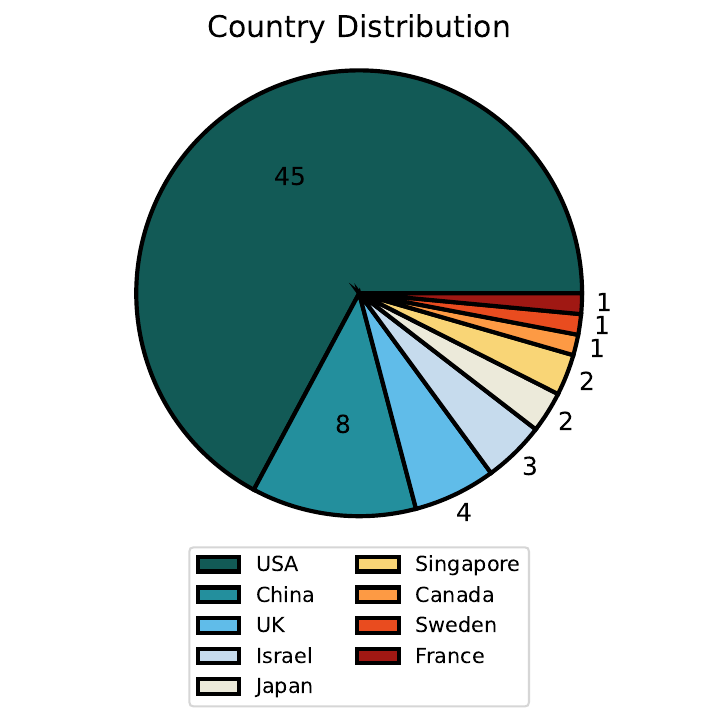}
    \caption{Most agentic systems are created by developers in the USA. In this figure, some developers' countries are counted multiple times due to producing multiple indexed systems. Google DeepMind is counted 3x, while OpenAI, National University of Singapore, UC Berkeley, and Stanford University are each counted 2x.}
    \label{fig:country}
\end{figure}

\textbf{Most indexed systems are created by developers located in the USA.} 
We considered the `developer country' of each agent to be the national location of either (a) the developer organization if the developer was a company or (b) the first author's first listed affiliation if the agent was created as part of an academic research collaboration. 
We plot the number of agents from each country in \Cref{fig:country}.
Of the 67 agents, 45 were created by developers in the USA.

\begin{figure}[h!]
    \centering
\includegraphics[width=0.85\linewidth]{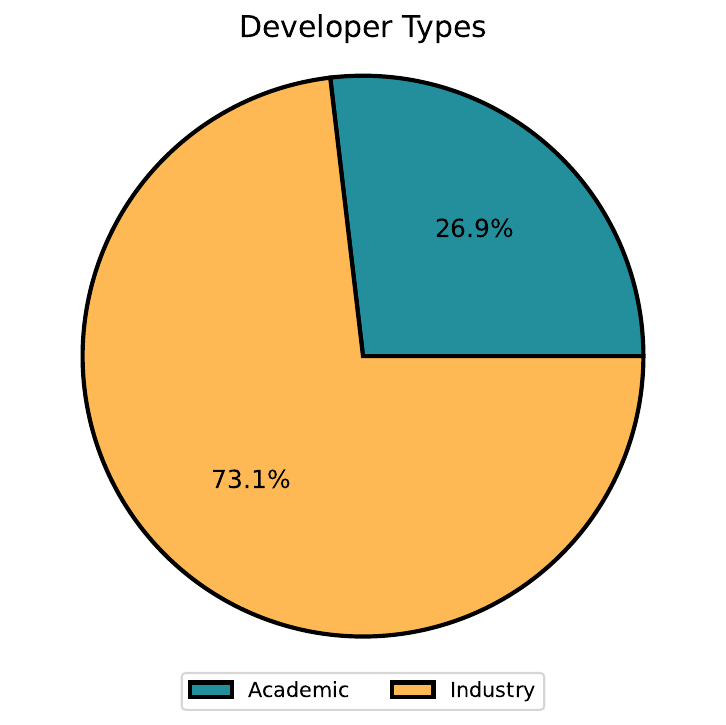}
    \caption{Most agentic systems are developed by companies.}
    \label{fig:industry_vs_academic}
\end{figure}

\textbf{While most agentic systems are developed by companies, a significant fraction are developed in academia.}
In \Cref{fig:industry_vs_academic}, we show the developers of agents broken down based on whether they are projects from academic labs or companies in industry. 
18 (26.9\%) are academic while 49 (73.1\%) are from companies.

\begin{figure}[h!]
    \centering
\includegraphics[width=\linewidth]{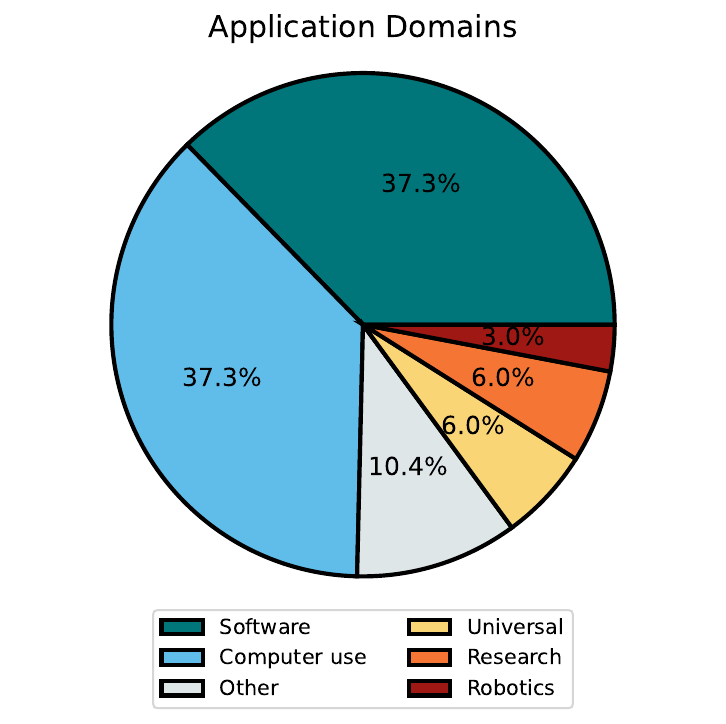}
    \caption{The majority of indexed systems specialize in software engineering and/or computer use.}
    \label{fig:application_domains}
\end{figure}

\textbf{The majority of indexed systems specialize in software engineering and/or computer use.} 
We divided the 67 systems into 6 categories:
\begin{itemize}[itemsep=6pt, topsep=6pt]
    \item \textit{Software}: agents that assist in coding and software engineering (e.g., \citealp{yang2024sweagent}). 
    \item \textit{Computer use}: agents designed to open-endedly interact with computer interfaces (e.g., \citealp{yoran2024assistantbench}) \citep{sager2025ai}.
    \item \textit{Universal}: agents designed to be a general-purpose reasoning engine (e.g., \citealp{openai2024o1preview}).
    \item \textit{Research}: agents designed to assist with scientific research (e.g., \citealp{lu2024ai}).
    \item \textit{Robotics}: agents designed for robotic control (e.g., \citealp{kim2024openvla}).
    \item \textit{Other}: systems that are designed for niche applications (e.g., \href{https://business.linkedin.com/talent-solutions/hiring-assistant}{LinkedIn Talent Agents}).
\end{itemize}
We plot the breakdown by domain in \Cref{fig:application_domains}.
50 of the 67 agents (74.6\%) specialize in either software engineering or computer use.
We also note that there exist many `agentic' systems for customer service, which do not meet our criteria for inclusion in the index. 
See \Cref{sec:methods} for discussion and examples.

\textbf{The majority of indexed systems have released code and/or documentation.}
Developers are relatively publicly forthcoming about details related to usage and capabilities.
In \Cref{fig:accessibility}, we show results: 
33 (49.3\%) release code, and 47 (70.1\%) release documentation. 
We also observed that systems developed as academic projects are released with a high degree of openness, with 16 of the 18 (88.8\%) releasing code. 

\textbf{There is limited publicly available information about safety testing and risk management practices.}
In contrast to the relatively high degree of openness that developers exercise around their systems' capabilities and usage, we find scant public information about safety policies, internal safety evaluations, and external safety evaluations. 
In \Cref{fig:safety}, we show that only 13 (19.4\%), 5 (7.5\%), and 6 (9\%) indexed systems have publicly available information on each of these, respectively. 
We also note that most of the systems that have undergone formal, publicly-reported safety testing are from a small number of large companies (e.g., Anthropic, Google DeepMind, OpenAI).

\section{Limitations and Concerns}
\label{sec:limitations}

\textbf{Defining agentic systems.} 
The term ``AI agent'' is contentious, as discussed in \Cref{sec:background}. 
In particular, the term has been criticized for inappropriately anthropomorphizing certain AI systems \citep{weidinger2022taxonomy, mitchell2021ai}, which could potentially lead to unrealistic expectations from, or over-reliance on, such systems \citep{gabriel2024ethics, manzini2024should}.
Recognizing this concern, we do not weigh in on this debate, advocate a particular definition of ``AI agent'', or propose alternative terminology.
Instead, we focus on empirically documenting a growing class of deployed AI systems that exhibit ``agentic'' characteristics (as described in \citet{chan2023harms}) and have a potential for significant impact. Through the index, we communicate our findings as plainly and openly as possible.

\textbf{Scope and timing of index.} 
The index is not a comprehensive or exhaustive database of all agentic systems or related resources, such as language models and development frameworks for building agentic systems.
The field of agentic AI is highly decentralized and poorly documented. 
Accordingly, there may also be systems that meet the selection criteria specified in \Cref{sec:methods} but do not appear in the index.
In particular, the index is likely to disproportionately document agentic systems that are publicly available or publicly released, compared with systems used internally within organizations (which, by definition, are not publicly accessible).
In addition, the index only includes systems described in the English language and includes relatively few systems from non-western developers. 
The index represents a snapshot in time on December 31, 2024 and does not include systems that were obsolete by this date or were released thereafter.
Moreover, while the agent cards in the index collect 33 fields of information, these are not exhaustive and exclude, for example, records of real-world safety incidents (to the extent such incidents have occurred).

\textbf{Incomplete or inaccurate information.} 
In total, the index contains over 2,200 fields of information reviewed by multiple authors.
Nonetheless, despite our best efforts to manually verify the completeness and accuracy of all agent cards, mistakes may have occurred.
In addition, the response rate of developers to our requests for feedback was 36\%. 
Accordingly, it is possible that some developers may, for example, have in place internal safety documents or practices that we could not discover from publicly available documentation, or were not informed about through correspondence. 
Recognizing these concerns, we have established a structured process for facilitating further corrections to the index. 
These can be submitted at \href{https://docs.google.com/forms/d/e/1FAIpQLScKdSu-jxDtBpbyfn5BebnBRYHYNJxMA6gij6RONVKefnR-cA/viewform?usp=header}{this link}.

\textbf{Promoting problematic practices.} 
The findings we present in \Cref{sec:findings}---particularly the lack of transparency around the safety features of agentic systems---could arguably promote problematic risk management practices.
For example, developers could choose to `game' an index like ours through perfunctory, selective disclosure of information recorded in the index \citep{krawiec2003cosmetic, marquis2016scrutiny}.
Due in part to this concern, we do not use this index to make developer scorecards. 
Instead, we see our findings as offering basic information to key stakeholders, including users, developers, auditors, and policymakers.
In doing so, we hope to lay the groundwork for more targeted assessments of impacts and risks from agentic systems in future work.

\section{Discussion and Future Work}
\label{sec:discussion_and_future_work}

\vspace{4pt}  

\textbf{The agentic AI ecosystem is difficult to document.} 
The extensive data collection process undertaken for the current paper (see \Cref{sec:methods}) sheds light on the significant challenges involved in documenting agentic AI systems. 
During this process, we encountered a diverse range of AI systems, across multiple domains, in different places in the research--product spectrum, and accompanied by varying levels of information and documentation.
The differences were often most stark when comparing systems developed in industry and systems developed in academia, the latter of which are typically simpler and more open.
On occasion, these features of the agentic AI ecosystem made it challenging to determine whether a particular system meets our criteria for inclusion in the index.
Most importantly, the fact that we ultimately produced an ``AI Agent Index'' should not be taken to suggest that this ecosystem lends itself to clean taxonomization and indexing (it does not). 
We expect these documentation challenges to persist for the foreseeable future.

\vspace{6pt}  

\textbf{Future documentation work should be appropriately scoped.} 
Our research design---including both the selection of information fields to be collected and the methods for collecting data---offers lessons for future attempts to document the agentic AI ecosystem.
From the outset, we sought to collect information on agentic systems that had been generally overlooked by previous survey papers and overviews of the field, such as the accessibility of documentation and code, information regarding red-teaming and safety policies, and the country of developers (see \Cref{sec:agent_card_components}).
Future documentation work can build on this approach, examining a broader range of technical, safety, and policy-relevant features of agentic AI systems.
To ensure tractability, we recommend that future work surveying the agent ecosystem be appropriately scoped either in breadth or depth.
For example, selection criteria could be revised to demand a high threshold for ``agency'' or anticipated societal impact.

\vfill
\break

\textbf{Documentation can inform governance and policy.}
Our findings (discussed in \Cref{sec:findings}) may inform the scope and methods of AI governance and policymaking:

\begin{itemize}[itemsep=2pt, topsep=2pt]
    \item The majority of indexed agentic systems were developed in industry, suggesting that governance interventions should consider the incentives of corporate developers (distinct from those of academic labs).
    \item Most indexed systems were developed by US-based organizations, indicating that governance efforts focused on US contexts could have more leverage than efforts in other countries or regions.
    \item The prominence of software engineering and computer-use agents suggests that policy researchers and practitioners should prioritize these domains when designing governance frameworks.
    \item Very few developers disclose information about safety or risk management, underscoring the importance of establishing transparency and disclosure mechanisms as a key first step in the governance of agentic systems.
\end{itemize}

To address knowledge and accountability gaps uncovered by our findings, policymakers could consider:
\begin{itemize}[itemsep=2pt, topsep=2pt]
    \item \textit{Structured bug bounties:} Incentivizing external red-teaming promotes the proactive discovery of vulnerabilities, adapting approaches used in cybersecurity.
    \item \textit{Systematic testing of agents:} Governance bodies and academic labs could coordinate risk assessments of agentic systems.
    \item \textit{Centralized oversight of indices:} Regulatory or standard-setting institutions could establish and maintain indices of agentic systems like this one.
    \item \textit{Integration with model registries:} Incorporate indices of agentic systems into broader registry frameworks \citep{mckernon2024ai}, ensuring unified reporting of agentic systems, common safety benchmarks, and clearer accountability mechanisms.
\end{itemize}

\section*{Impact Statement}
This work was undertaken to improve our collective understanding of the emerging field of agentic AI. 
Its contributions revolve around the compilation and analysis of publicly available information, supplemented by correspondence with developers.
In \Cref{sec:limitations}, we discuss how transparency standards can be `gamed,' and note that this was one reason that we did not score developers using the index.
Taken together, we hope the methodology and findings introduced by the AI Agent Index inform progress toward better risk management practices and governance frameworks for agentic AI systems. 

\newpage

\section*{Acknowledgments}

We are thankful to Alan Chan, Atoosa Kasirzadeh, Laker Newhouse, Gabe Mukobi, Rishi Bommasani, Peter Cihon, Merlin Stein, Greg Leppert, Jack Cushman, and Seth Lazar for discussions and feedback.

\bibliography{bibliography}
\bibliographystyle{icml2025}

\newpage
\appendix
\onecolumn

\section{Sample Agent Card} \label{app:sample_agent_card}

Here, we provide a sample agent card for Microsoft's Magentic One \citep{fourney2024magentic}.
We selected it based on its recency, degree of documentation, openness, generality, and noteworthy performance.
No authors have conflicts of interest related to Microsoft or Magentic One, and this example selection was made without correspondence with Microsoft. 
Including Magentic One's agent card as an example is not an endorsement of the system or developer. 

\subsection*{Magentic One}

\begin{enumerate}

    \item \textbf{Basic information}
    \begin{itemize}
        \item Website: \href{https://www.microsoft.com/en-us/research/publication/magentic-one-a-generalist-multi-agent-system-for-solving-complex-tasks/}{https://www.microsoft.com/en-us/research/publication/magentic-one-a-generalist-multi-agent-system-for-solving-complex-tasks/}
        \item Short description: A multiagent system introduced by Microsoft with general capabilities.
        \item Intended uses: \textit{What does the developer state that the system is intended for?} It is used for ``ad-hoc, open-ended tasks such as browsing the web and interacting with web-based applications, handling files, and writing and executing Python code'' [\href{https://www.microsoft.com/en-us/research/publication/magentic-one-a-generalist-multi-agent-system-for-solving-complex-tasks/}{source}].
        \item Date(s) deployed: Announced November 4, 2023 [\href{https://web.archive.org/web/20241231233125/https://www.microsoft.com/en-us/research/articles/magentic-one-a-generalist-multi-agent-system-for-solving-complex-tasks/}{source}].
    \end{itemize}

    \item \textbf{Developer}
    \begin{itemize}
        \item Website: \href{https://web.archive.org/web/20241231232226/https://www.microsoft.com/en-us/}{https://web.archive.org/web/20241231232226/https://www.microsoft.com/en-us/}
        \item Legal name: Microsoft Corporation [\href{https://web.archive.org/web/20250105103050/https://www.microsoft.com/en-US/servicesagreement/}{source}].
        \item Entity type: Corporation [\href{https://web.archive.org/web/20250105103050/https://www.microsoft.com/en-US/servicesagreement/}{source}].
        \item Country (location of developer or first author's first affiliation): Incorporation: Washington, USA (Microsoft Corporation (2357303)) [\href{https://www.google.com/url?q=https://icis.corp.delaware.gov/eCorp/EntitySearch/NameSearch.aspx&sa=D&source=editors&ust=1730995934133445&usg=AOvVaw0-1z61YlnmkW_0dL2g9R2-}{source}]. Registration: Delaware, USA. HQ: Washington, USA [\href{https://plus.lexis.com/document?pdmfid=1530671&pddocfullpath=%2Fshared%2Fdocument%2Fcompany-financial%2Furn%3AcontentItem%3A61CW-WMC3-HFSB-30SB-00000-00&pdcontentcomponentid=428957&pdislparesultsdocument=false&prid=bcee593f-5fe8-4b34-8610-baaf89722c00&crid=e0ef80ad-5758-4d8d-aa23-2924f95722fb&pdisdocsliderrequired=true&pdpeersearchid=d8a521e8-1774-495b-8601-d47ca2a084a7-1&ecomp=undefined&earg=sr2#/document/93b2613c-d815-4baf-b071-eee7dc030114}{source}].
        \item Safety policies: \textit{What safety and/or responsibility policies are in place?} Model evaluations and red teaming; model reporting and information sharing; security controls [\href{https://web.archive.org/web/20241223045258/https://blogs.microsoft.com/on-the-issues/2023/10/26/microsofts-ai-safety-policies/}{source}]. Microsoft’s safety policies are described online [\href{https://web.archive.org/web/20241222064226/https://www.microsoft.com/en-us/ai/responsible-ai}{source}].
    \end{itemize}

    \item \textbf{System components}
    \begin{itemize}
        \item Backend model: \textit{What model(s) are used to power the system?} The default model used is gpt-4o-2024-05-13, but they also experiment with using OpenAI o1 [\href{https://www.microsoft.com/en-us/research/publication/magentic-one-a-generalist-multi-agent-system-for-solving-complex-tasks/}{source}].
        \item Publicly available model specification: \textit{Is there formal documentation on the system’s intended uses and how it is designed to behave in them?} Available [\href{https://web.archive.org/web/20241228060554/https://www.microsoft.com/en-us/ai/principles-and-approach}{source}].
        \item Reasoning, planning, and memory implementation: \textit{How does the system ‘think’?} The system contains multiple subagents that work together to solve problems. Things are controlled at a high level by the “Orchestrator” agent and executed by the “WebSurfer,” FileSurfer,” “Coder,” and “ComputerTerminal” agents [\href{https://www.microsoft.com/en-us/research/publication/magentic-one-a-generalist-multi-agent-system-for-solving-complex-tasks/}{source}].
        \item Observation space: \textit{What is the system able to observe while `thinking’?} It has full access to a filesystem and web browser.
        \item Action space/tools: \textit{What direct actions can the system take?} It is able to surf (including posting) on the web, execute file system commands, and write/execute code.
        \item User interface: \textit{How do users interact with the system?} Users can configure and experiment with it using the AutoGen package [\href{https://web.archive.org/web/20241219131707/https://github.com/microsoft/autogen/tree/main/python/packages/autogen-magentic-one}{source}].
        \item Development cost and compute: \textit{What is known about the development costs?} Unknown.
    \end{itemize}

    \item \textbf{Guardrails and oversight}
    \begin{itemize}
        \item Accessibility of components:
        \begin{itemize}
            \item Weights: \textit{Are model parameters available?} N/A; backends various models.
            \item Data: \textit{Is data available?} N/A; backends various models.
            \item Code: \textit{Is code available?} Available on GitHub as part of Microsoft’s AutoGen project [\href{https://web.archive.org/web/20250105175141/https://github.com/microsoft/autogen}{source}].
            \item Scaffolding: \textit{Is system scaffolding available?} Available [\href{https://web.archive.org/web/20241219131707/https://github.com/microsoft/autogen/tree/main/python/packages/autogen-magentic-one}{source}].
            \item Documentation: \textit{Is documentation available?} Available on GitHub [\href{https://web.archive.org/web/20241219131707/https://github.com/microsoft/autogen/tree/main/python/packages/autogen-magentic-one}{source}], see also the technical report [\href{https://www.microsoft.com/en-us/research/publication/magentic-one-a-generalist-multi-agent-system-for-solving-complex-tasks/}{source}].
        \end{itemize}
        \item Controls and guardrails: \textit{What notable methods are used to protect against harmful actions?} The developers recommend using containers, virtual environments, log monitoring, human oversight, access limitations, and data safeguards.
        \item Customer and usage restrictions: \textit{Are there know-your-customer measures or other restrictions on customers?} None.
        \item Monitoring and shutdown procedures: \textit{Are there any notable methods or protocols that allow for the system to be shut down if it is observed to behave harmfully?} Logs are kept while the system runs.
    \end{itemize}

    \item \textbf{Evaluations}
    \begin{itemize}
        \item Notable benchmark evaluations (e.g., on SWE-Bench Verified): GAIA (38\%), AssistantBench (27.7\%), and WebArena (32.8\%) [\href{https://www.microsoft.com/en-us/research/publication/magentic-one-a-generalist-multi-agent-system-for-solving-complex-tasks/}{source}].
        \item Bespoke testing (e.g., demos): None.
        \item Safety: \textit{Have safety evaluations been conducted by the developers? What were the results?} They report on ad-hoc evaluations of failures and safety concerns in the technical report [\href{https://www.microsoft.com/en-us/research/publication/magentic-one-a-generalist-multi-agent-system-for-solving-complex-tasks/}{source}]. The developers claim: “We performed testing for Responsible AI harm e.g., cross-domain prompt injection and all tests returned the expected results with no signs of jailbreak” [\href{https://web.archive.org/web/20250102132332/https://github.com/microsoft/autogen/blob/main/TRANSPARENCY_FAQS.md}{source}].
        \item Publicly reported external red-teaming or comparable auditing:
        \begin{itemize}
            \item Personnel: \textit{Who were the red-teamers/auditors?} None.
            \item Scope, scale, access, and methods: \textit{What access did red-teamers/auditors have and what actions did they take?} None.
            \item Findings: \textit{What did the red-teamers/auditors conclude?} None.
        \end{itemize}
    \end{itemize}

    \item \textbf{Ecosystem}
    \begin{itemize}
        \item Interoperability with other systems: \textit{What tools or integrations are available?} It was not explicitly designed to interoperate with any particular systems other than the web browser and filesystem. But it presumably could integrate with others with little configuration.
        \item Usage statistics and patterns: \textit{Are there any notable observations about usage?} Microsoft AutoGen has 36.9k stars and 5.3k forks [\href{https://github.com/microsoft/autogen/tree/gaia_multiagent_v01_march_1st}{source}].
    \end{itemize}

    \item \textbf{Additional notes:} None.
    
\end{enumerate}

\end{document}